\documentclass[twoside,a4paper,11pt]{sea10}
\usepackage{graphicx}
\usepackage{hyperref}
\usepackage{movie15}
\usepackage{amssymb} 
\begin{document}
\pagenumbering{arabic}
\setcounter{page}{280}
\pagestyle{myheadings}
\thispagestyle{empty}
{\flushleft\includegraphics[width=\textwidth,bb=58 650 590 680]{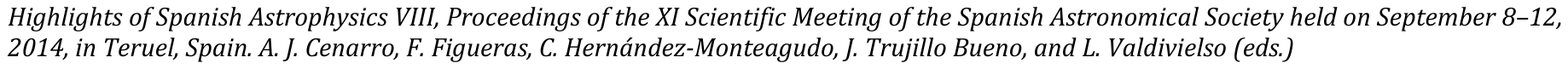}}
\vspace*{0.2cm}
\begin{flushleft}
{\bf {\LARGE
%
Which galaxy mass estimator can we trust?
%
}\\
\vspace*{1cm}
%
Luis Peralta de Arriba$^{1,2}$,
Marc Balcells$^{3,1,2}$,
Jes\'us Falc\'on-Barroso$^{1,2}$,
and
Ignacio Trujillo$^{1,2}$
%
}\\
\vspace*{0.5cm}
%
$^{1}$
Instituto de Astrof\'{\i}sica de Canarias, E-38200 La Laguna, Tenerife, Spain\\
$^{2}$
Universidad de La Laguna, Departamento de Astrof\'{\i}sica, E-38206 La Laguna, Tenerife, Spain\\
$^{3}$
Isaac Newton Group of Telescopes, E-38700 Santa Cruz de La Palma, La Palma, Spain
%
\end{flushleft}
%
\markboth{
Which galaxy mass estimator can we trust?
}{ 
%
L. Peralta de Arriba et al.
%
}
\thispagestyle{plain}
\vspace*{0.4cm}
\begin{minipage}[l]{0.09\textwidth}
\ 
\end{minipage}
\begin{minipage}[r]{0.9\textwidth}
\vspace{1cm}
\section*{Abstract}{\small
%
We address the problem that dynamical masses of high-redshift massive galaxies, derived using virial scaling, often come out lower than stellar masses inferred from population fitting to multi-band photometry. We compare dynamical and stellar masses for various samples spanning ranges of mass, compactness and redshift, including the SDSS. The discrepancy between dynamical and stellar masses occurs both at low and high redshifts, and systematically increases with galaxy compactness. Because it is unlikely that stellar masses show systematic errors with galaxy compactness, the correlation of mass discrepancy with compactness points to errors in the dynamical mass estimates which assume homology with massive, nearby ellipticals. We quantify the deviations from homology and propose specific non-virial scaling of dynamical mass with effective radius and velocity dispersion.
%
\normalsize}
\end{minipage}
%
%
%
\section{Introduction to the discrepancy between dynamical and stellar masses} \label{intro}
As soon as measurements of internal velocity dispersions of high-$z$ galaxies became available, several authors (e.g. \cite{2012MNRAS.423..632F,2011ApJ...738L..22M,2010ApJ...709L..58S,2014ApJ...780..134S}) reported that implied dynamical masses ($M_\mathrm{dyn}$) were lower than stellar masses ($M_\star$). This is unphysical, as dynamical masses weigh stars plus dark matter and gas, while stellar masses measure stars only.

Stellar masses are derived from stellar population fitting, while dynamical masses are estimated assuming virial equilibrium and homology. These mass estimators are the only ones available for high-$z$ galaxies. We summarize some features of both mass estimators in the next subsections.

\subsection{Stellar mass from stellar populations} \label{subsec:stellar_mass}
This method consists in fitting spectra from stellar population synthesis models to a galaxy spectral energy distribution. This method is powerful because it can be applied to broadband photometric data. Its main uncertainties come from incomplete knowledge of the stellar initial mass function (IMF), the star formation history and the dust attenuation. In addition, late phases of stellar evolution such as the asymptotic giant branch are difficult to model, and introduce important uncertainties in the stellar mass.

\subsection{Dynamical mass from virial theorem with homology} \label{subsec:dyn_mass}
The virial theorem connects the kinetic and potential energies and, hence, links internal velocities, galaxy size and dynamical mass. Assuming that early-type galaxies (ETGs) are homologous systems, i.e. they have similar dynamical structures (they share similar density, kinematic and luminosity profiles), one can use this theorem to estimate dynamical masses. In particular, the above two hypotheses lead to
\begin{equation} \label{eq:virial}
M_\mathrm{dyn} = K \frac{\sigma_\mathrm{e}^2 r_\mathrm{e}}{G},
\end{equation}
where $K$ is a constant for all ETG, $r_\mathrm{e}$ the effective radius, $\sigma_\mathrm{e}$ the velocity dispersion within the effective radius and $G$ the gravitational constant.

The validity of both hypotheses was tested in normal-sized ETGs in the local Universe as part of the SAURON project \cite{2006MNRAS.366.1126C}. Their galaxies were compatible with the assumption virial equilibrium and homology, and gave a calibration for the previous equation ($K=5.0$).

Dynamical masses require velocity dispersion measurements, which are costly in telescope time, and, hence, have only become available in recent years.

\section{Characterization of the discrepancy} \label{characterization}
In this section we explain the characterization of the discrepancy performed by \cite{2014MNRAS.440.1634P}. Specifically, we describe the data of this work and we explain briefly its results.

\subsection{Data description}
All the galaxies used by \cite{2014MNRAS.440.1634P} satisfy the next points:
\begin{itemize}
\item They are early-type, i.e. they have S\'ersic indices $n > 2.5$.
\item They are massive, i.e. they satisfy $M_\star \gtrsim 10^{11} \ \mathrm{M_\odot}$.
\item Their redshift range is $0.0 \lesssim z \lesssim 2.5$.
\item They come from several data sources, including a sample of $\sim$54000 at $z \sim 0.08$ from SDSS NYU (\cite{2007AJ....133..734B,2005AJ....129.2562B}).
\end{itemize}

\subsection{Does the $M_\star / M_\mathrm{dyn}$ ratio depend on redshift?}
\begin{figure}
\center
\includegraphics[width=75mm]{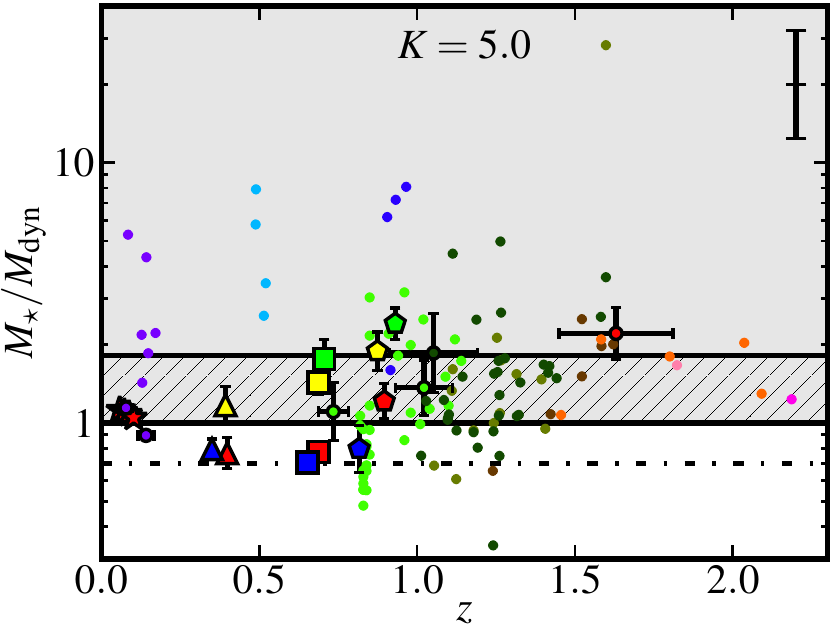}
\caption{\label{fig1} $M_\star / M_\mathrm{dyn}$ ratio versus redshift. Figure from \cite{2014MNRAS.440.1634P}.}
\end{figure}

Figure~\ref{fig1} shows that there is not a clear trend between the $M_\star / M_\mathrm{dyn}$ ratio and redshift. However, we can check that $M_\star / M_\mathrm{dyn} > 1$ is detected more often at high redshift. This suggests that the discrepancy could be connected with galaxy compactness, and then the strong size evolution of massive galaxies (e.g. \cite{2007MNRAS.382..109T}) would explain why the discrepancy is reported more often at high redshift. This possibility will be addressed in the next subsection.

\subsection{Does the $M_\star / M_\mathrm{dyn}$ ratio depend on galaxy compactness?} \label{subsec:compactness}
\begin{figure}
\center
\includegraphics[width=75mm]{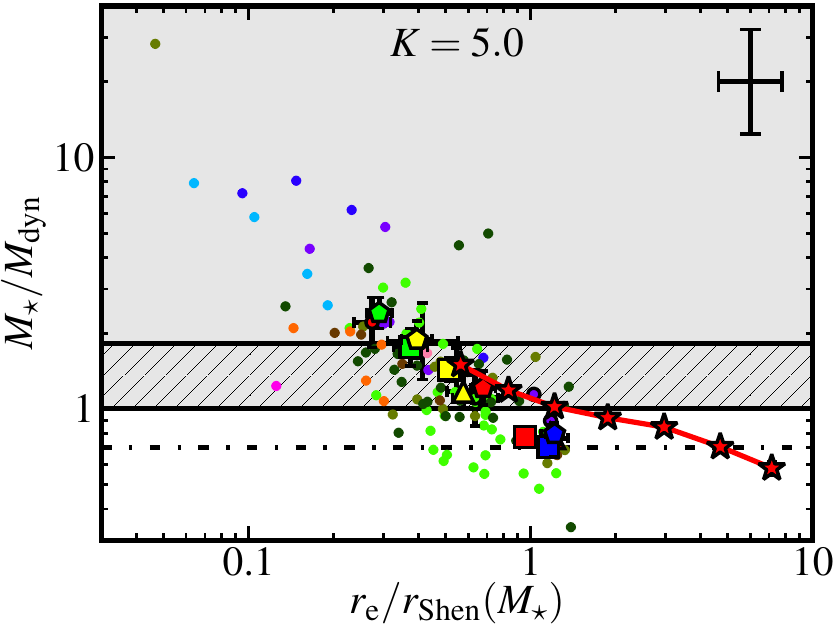} ~
\includegraphics[width=75mm]{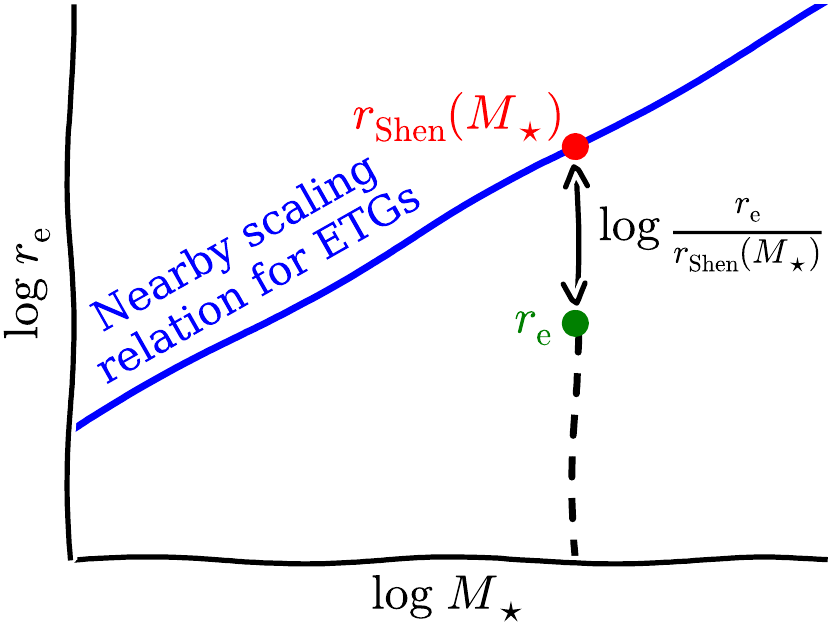}
\caption{\label{fig2} Left panel shows $M_\star / M_\mathrm{dyn}$ ratio versus the compactness indicator $r_\mathrm{e} / r_\mathrm{Shen}(M_\star)$. The hatched region of left panel indicates the area where the discrepancy can be solved tuning the IMF. Right panel is a schematic illustration which shows the geometrical interpretation of the compactness indicator $r_\mathrm{e} / r_\mathrm{Shen}(M_\star)$: it is the distance of a galaxy to the nearby scaling relationship for ETGs determined by \cite{2003MNRAS.343..978S}. Left panel from \cite{2014MNRAS.440.1634P}.}
\end{figure}

The left panel of Fig.~\ref{fig2} shows that there is a clear trend between the $M_\star / M_\mathrm{dyn}$ ratio and the compactness indicator $r_\mathrm{e} / r_\mathrm{Shen}(M_\star)$. The compactness indicator $r_\mathrm{e} / r_\mathrm{Shen}(M_\star)$ of a galaxy is defined as the ratio between its radius and the radius that it would have if it followed the nearby scaling relationship for ETGs determined by \cite{2003MNRAS.343..978S}. The right panel of Fig.~\ref{fig2} illustrates the geometric interpretation of this compactness indicator: it is the distance of a galaxy to the nearby scaling relationship for ETGs.

As the IMF is the main uncertainty in the estimation of stellar mass, one could think that tuning this parameter the discrepancy perhaps could be solved. The hatched region of left panel of Fig.~\ref{fig2} indicates the area where the discrepancy, $M_\star / M_\mathrm{dyn} > 1$, can be solved by tuning the IMF. Figure~\ref{fig2} shows that many galaxies lie in the forbidden region of the diagram even after tuning the IMF.

\begin{figure}
\center
\includegraphics{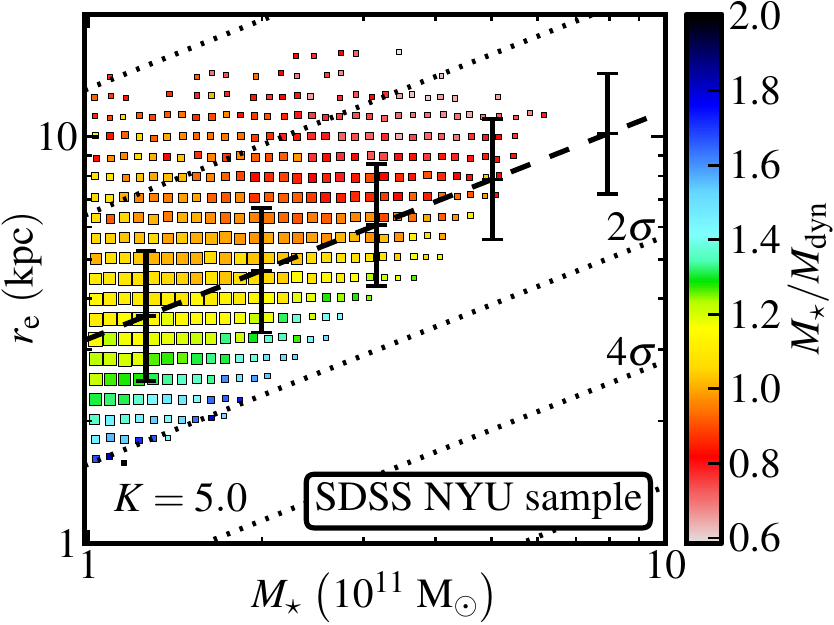}
\caption{\label{fig3} Dependency of $M_\star / M_\mathrm{dyn}$ ratio on the stellar mass--size plane for the SDSS NYU sample. Dashed line is the the nearby scaling relationship for ETGs determined by \cite{2003MNRAS.343..978S}. Figure from \cite{2014MNRAS.440.1634P}.}
\end{figure}

Figure~\ref{fig3} shows that the trend between $M_\star / M_\mathrm{dyn}$ ratio and the compactness indicator $r_\mathrm{e} / r_\mathrm{Shen}(M_\star)$ is also present in the nearby Universe.

\section{A possible solution to $M_\star / M_\mathrm{dyn} > 1$} \label{solution}
In this section we explain a possible solution to the discrepancy between masses proposed by \cite{2014MNRAS.440.1634P}.

Recalling the description of both mass estimators in Sect.~\ref{subsec:stellar_mass} and \ref{subsec:dyn_mass}, and given that changing the IMF does not solve the discrepancy (see Sect.~\ref{subsec:compactness}), we investigate whether relaxing the homology assumption can bring stellar and dynamical masses to be consistent with each other.

Assuming that homology is not satisfied on any ETG, a solution to the discrepancy between stellar and dynamical masses is to make the $K$ parameter from Eq.~(\ref{eq:virial}) depend on galaxy compactness. Doing this it is found that $K$ should satisfy the next equation:
\begin{equation} \label{eq:k}
K \sim 6.0 \,  \left( \frac{r_\mathrm{e}}{3.185 \  \mathrm{kpc}} \right)^{-0.81} \left( \frac{M_\star}{10^{11} \  \mathrm{M_\odot}} \right)^{0.45}.
\end{equation}

\begin{figure}
\center
\includegraphics{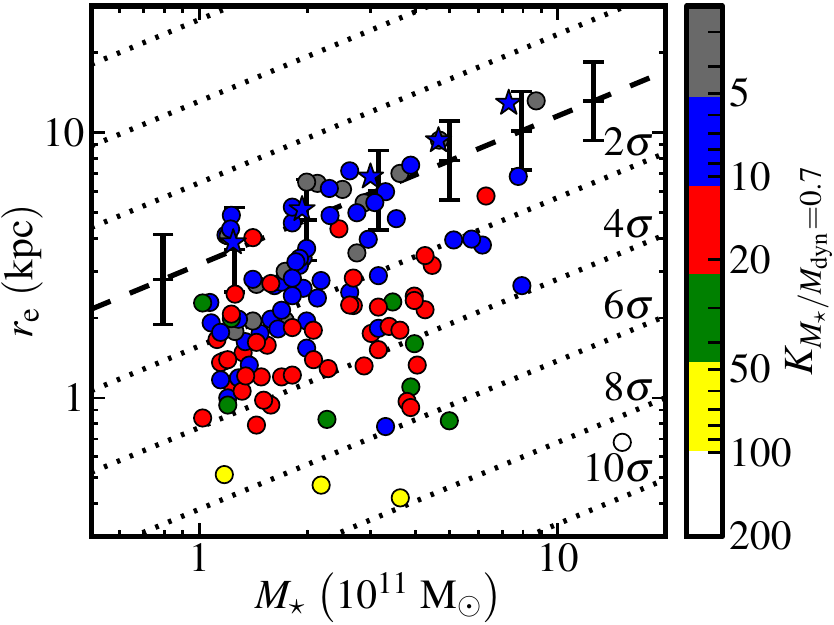}
\caption{\label{fig4} Variation of $K$ to solve mass discrepancies depending on the stellar mass--size plane position. Dashed line is the nearby scaling relationship for ETGs determined by \cite{2003MNRAS.343..978S}. Figure from \cite{2014MNRAS.440.1634P}.}
\end{figure}

Figure~\ref{fig4} shows the results of applying the Eq.~(\ref{eq:k}). It is worth noting that this equation implies $K$ values which are often an order of magnitude higher than the value measured on the massive, nearby, normal-sized ellipticals.

Eq.~(\ref{eq:k}) also implies the next non-virial scaling of dynamical mass with effective radius and velocity dispersion:
\begin{equation} \label{eq:non_virial}
M_\mathrm{dyn} \sim \left( \frac{\sigma_\mathrm{e}}{200 \  \mathrm{km \  s^{-1}}} \right)^{3.6} \left( \frac{r_\mathrm{e}}{3 \  \mathrm{kpc}} \right)^{0.35} 2.1 \times 10^{11} \  \mathrm{M_\odot}.
\end{equation}

\section{Takeaway messages} \label{messages}
The takeaway messages of the work by \cite{2014MNRAS.440.1634P} are:
\begin{itemize}
\item There is a discrepancy between dynamical and stellar masses:

Several authors (e.g. \cite{2012MNRAS.423..632F,2011ApJ...738L..22M,2010ApJ...709L..58S,2014ApJ...780..134S}) have reported that they find stellar masses greater than virial masses at low and high redshift.

\item This discrepancy scales with galaxy compactness:

The discrepancy between dynamical and stellar masses increases as galaxies depart from the nearby scaling relationship between stellar mass and size for ellipticals.

\item Non-homology can solve the discrepancy:

As the validity of the homology assumption has been only tested in the nearby scaling relationship between stellar mass and size, the most likely solution seems that the dynamical structure of ellipticals depends on galaxy compactness.

\end{itemize}

\section*{Acknowledgments}   
%
{\small LPdA was supported by the Spanish Ministry of Science and Innovation. LPdA would like to thank M. Ram\'on-P\'erez for her comments that helped to improve the presentation of our results. This work has been supported by the Programa Nacional de Astronom\'{\i}a y Astrof\'{\i}sica of the Spanish Ministry of Science and Innovation under grants AYA2009-11137 and AYA2010-21322-C03-02. \normalsize}

%

%
\end{document}